\documentclass[aip,preprint]{revtex4-1}
\usepackage{graphicx}

\draft % marks overfull lines with a black rule on the right

\begin{document}

% Use the \preprint command to place your local institutional report number 
% on the title page in preprint mode.
% Multiple \preprint commands are allowed.
%\preprint{}

\title{Gafchromic EBT2 film dosimetry in reflection mode with a novel plan-based calibration method} 

% repeat the \author .. \affiliation  etc. as needed
% \email, \thanks, \homepage, \altaffiliation all apply to the current author.
% Explanatory text should go in the []'s, 
% actual e-mail address or url should go in the {}'s for \email and \homepage.
% Please use the appropriate macro for the type of information
% \affiliation command applies to all authors since the last \affiliation command. 
% The \affiliation command should follow the other information.

\author{I. M{\'e}ndez}
\email[]{nmendez@onko-i.si}
\author{V. Hartman}
\author{R. Hudej}
\author{A. Strojnik}
\author{B. Casar}
\affiliation{Department of Medical Physics, Institute of Oncology Ljubljana, Zalo\v{s}ka cesta 2, Ljubljana 1000, Slovenia}

% Collaboration name, if desired (requires use of superscriptaddress option in \documentclass). 
% \noaffiliation is required (may also be used with the \author command).
%\collaboration{}
%\noaffiliation

%\date{\today}

\begin{abstract}
\textbf{Purpose:}
A dosimetric system formed by Gafchromic EBT2 radiochromic film and Epson Expression 10000XL flatbed scanner was commissioned for dosimetry. In this paper, several open questions concerning the commissioning of radiochromic films for dosimetry were addressed: a) is it possible to employ this dosimetric system in reflection mode; b) if so, can the methods used in transmission mode also be used in reflection mode; c) is it possible to obtain accurate absolute dose measurements with Gafchromic EBT2 films; d) which calibration method should be followed; e) which calibration models should be used; f) does three-color channel dosimetry offer a significant improvement over single channel dosimetry. The purpose of this paper is to help clarify these questions.

\noindent\textbf{Methods:}
In this study, films were scanned in reflection mode, the effect of surrounding film was evaluated and the feasibility of EBT2 film dosimetry in reflection mode was studied. EBT2's response homogeneity has been reported to lead to excessive dose uncertainties. To overcome this problem, a new plan-­based calibration method was implemented. Plan­-based calibration can use every pixel and each of the three color channels of the scanned film to obtain the parameters of the calibration model. A model selection analysis was conducted to select lateral correction and sensitometric curve models. The commonly used calibration with fragments was compared with red-­channel plan-­based calibration and with three-­channel plan-­based calibration.

\noindent\textbf{Results:}
No effect of surrounding film was found in this study. The film response inhomogeneity in EBT2 films was found to be important not only due to differences in the fog, but also due to differences in sensitivity. The best results for lateral corrections were obtained using absolute corrections independent of the dose. With respect to the sensitometric curves, an empirical polynomial fit of order 4 was found to obtain results equivalent to a gamma-distributed single hit model based on physical assumptions. Three-channel dosimetry was found to be substantially superior to red-channel dosimetry.

\noindent\textbf{Conclusions:}
Reflection mode with Gafchromic EBT2 radiochromic film was found to be a viable alternative to transmission mode. The same methods that are used in transmission mode can be followed in reflection mode. A novel plan-based method was developed for calibration and multichannel dosimetry. 
This novel method offers increased robustness against film response inhomogeneities and reduces considerably the time required for calibration.

\end{abstract}

% insert suggested PACS numbers in braces on next line

\maketitle %\maketitle must follow title, authors, abstract and \pacs

\section{Introduction}

Radiochromic films present weak energy dependence, high spatial resolution, and near water equivalence. This makes them appropriate for measurements whenever nonequilibrium conditions exist, in fields with high dose gradients and in tissue heterogeneities: particularly for advanced radiotherapy techniques such as intensity modulated radiotherapy (IMRT) or stereotactic radiosurgery (SRS). Nevertheless, some questions remain open when radiochromic films are commissioned for dosimetry. The purpose of this paper is to help clarify these questions.

In this research, a dosimetric system formed by Gafchromic EBT2 radiochromic film (International Specialty Products, Wayne, NJ) and Epson Expression 10000XL flatbed scanner (Seiko Epson Corporation, Nagano, Japan) was commissioned for dosimetry. The Epson Expression 10000XL can scan either in reflection mode or in transmission mode, the later with a transparency adapter purchased separately. Then, the first question to answer is whether it is possible to employ this dosimetric system in reflection mode. If so, it should be investigated whether the methods used in transmission mode can also be used in reflection mode. Radiochromic film dosimetry has been previously developed in reflection mode\cite{Alva:2002,kalef-ezra:2008}. Nevertheless, to the authors' knowledge, the only analysis of dosimetry in reflection mode using Gafchromic EBT films (namely EBT2 films) was performed by Richley \emph{et al} (2010)\cite{Richley:2010}, who reported an effect of surrounding film that makes it impossible to use the same protocols that are used in transmission mode dosimetry also in reflection mode. In this study films were scanned in reflection mode, the effect of surrounding film was evaluated and the feasibility of EBT2 film dosimetry in reflection mode was studied.

Gafchromic EBT film was replaced by EBT2 film in 2009. EBT film has been extensively commissioned and found to be reliable for dose measurements \cite{fiandra:2006, Fuss:2007, Paelinck:2007, battum:2008, Martisikova:2008, menegotti:2008}. However, EBT2 properties have been studied in several papers\cite{arjomandy:2010, devic:2010,andres:2010, Richley:2010} and doubts have been cast on its response homogeneity \citep{hartmann:2010, Kairn:2010}, which has been reported to lead to excessive dose uncertainties. A question arises whether it is possible to obtain accurate absolute dose measurements with Gafchromic EBT2 films. The answer to this question depends on the calibration method employed. In the literature, the most frequent calibration method uses fragments irradiated with different doses and scanned in different positions over the scanner\cite{bouchard:2009, devic:2005, Martisikova:2008, Fuss:2007}. Another faster and accurate method was proposed by Menegotti \emph{et al} (2008)\cite{menegotti:2008} who used single film exposure. A weakness of this method was the fact that only six levels of uniform dose placed in stripes were used to parametrize the sensitometric curve. This could reduce the accuracy of the sensitometric curve\citep{bouchard:2009}, especially when film response is inhomogeneous. To offer increased robustness against film response inhomogeneities, while being faster than the conventional fragment­ based method, a novel plan-based calibration method was developed in this work. Independently of the calibration method, a suitable calibration model (consisting of lateral correction and sensitometric curve models) should be chosen. In this paper, a model selection analysis based on maximum likelihood estimation was performed to select the calibration model. The last question addressed in this paper is whether three-channel dosimetry offers a significant improvement over one channel dosimetry. Until recently\cite{AMicke:2011, mccaw:2011, mayer:2012}, only one color channel has been commonly used for radiochromic film dosimetry. The red channel has been chosen because it has been found to provide the greatest sensitivity at lower doses \cite{Richley:2010}. Three-channel dosimetry using the weighted mean of the channels was developed both for calibration with fragments and plan-based calibration, and their results were compared.

\section{Methods and materials}

\subsection{Dosimetric system}

Gafchromic EBT2 films with dimensions 8 ${\rm inch}$ $\times$ 10 ${\rm inch}$ were used in this work. They were handled following recommendations outlined in the AAPM TG-55 report \cite{aapm:55}. When smaller film pieces were required, films were divided into fragments with dimensions:  6.4 ${\rm cm}$ $\times$ 6.8 ${\rm cm}$. The lot used was A04141003BB, except in the analysis of intralot and interlot variations, which included also lot A03171101A. 

Films were digitized with an Epson Expression 10000XL flatbed scanner. This device is a 48-bit color scanner equipped with a linear charge-coupled device (CCD) as optical sensor, a xenon lamp as the light source and which can scan either in transmission mode (if a transparency adapter is acquired) or in reflection mode. In this work, each film was scanned in reflection mode.

\subsection{Irradiation procedure}

The films employed in preliminary tests and calibrations were irradiated in a 12$\times$30$\times$30 ${\rm cm^3}$ Plastic Water\texttrademark phantom (Computerized Imaging Reference Systems Inc. Norfolk, VA, USA) with a 6 MV photon beam from a Novalis Tx accelerator (Varian, Palo Alto, CA, USA). Films were centered on the beam axis at a depth of 6 cm using SAD (source-axis distance) set-up.  

The films employed in verification tests were irradiated in a CIRS Thorax phantom (Model 002LFC) with a 6 MV photon beam from a Varian Unique\texttrademark accelerator (Varian, Palo Alto, CA, USA). The CIRS Thorax phantom represents an average human torso, both in dimensions and structure. Its body is made of plastic water and it includes tissue heterogeneities corresponding to lung and bone. The films were placed with an offset of 1.5 cm from the beam axis to avoid the film and the beam axis being in the same plane \cite{Kunzler:2009}. The phantom was set-up at SAD.

\subsection{Scanning protocol}

Before acquisitions, the scanner was allowed to warm up for 30 min.  
Each film or film fragment was scanned in reflection mode and portrait orientation, 24 h after exposure (within a time window of less than 1 h)\cite{devic:2010} and centered on the scanner with a black opaque cardboard frame.

Films were digitized using the associated software Epson Scan v.3.0. Images were acquired in "professional mode" with the image type set to 48-bit RGB (16 bit per channel), depending on the test the resolution was either 72 dpi or 150 dpi (0.35 mm/px or 0.17 mm/px) and the image processing tools were turned off. Data were saved as TIFF (tagged image file format) files. 

Five consecutive scans were made for each film. The warm-up effect of the scanner lamp due to multiple scans\cite{Paelinck:2007, Martisikova:2008} was studied during the commissioning. The first scan was found to be markedly different from the last four scans and was therefore discarded. The resulting image was the average of the last four. Images were analyzed with the  open-source software ImageJ v.1.44o (National Institutes of Health, USA).

\subsection{Preliminary tests}

\subsubsection{Effect of surrounding film}

Richley \emph{et al} (2010)\citep{Richley:2010} reported that the pixel value (PV) of a region of interest (ROI) was dependent on the PV of the surrounding film when scanned in reflection mode with Epson Expression 10000XL. This effect would implicate a serious disadvantage for dosimetry in reflection mode compared to transmission mode, requiring more tests and worsening the uncertainties of the dosimetric system.

To investigate this effect, seven fragments of a film were used. Six of them were irradiated with 1, 2, 3, 4, 5 and 6 Gy, respectively. The nonirradiated film was centered on the scanner and one of the irradiated pieces was positioned next to it along the {\it x} axis of the scanner ({\it i.e.}, parallel to the CCD array). Images were acquired with resolution of 72 dpi, and a 50$\times$50 px ROI was measured at the center of the nonirradiated film. The process was repeated with each irradiated fragment.       

\subsubsection{Film response homogeneity}

To examine the EBT2 film response homogeneity a film was cut into 12 fragments. Before irradiation, and 24h after being irradiated with 2 Gy, each fragment was centered on the scanner, and a 100$\times$100 px ROI (3.5$\times$3.5 ${\rm cm^2}$) was measured at the center of the fragment.  

\subsection{Calibration}

Subtracting the optical density (OD) of a film before irradiation from the OD after irradiation improves the accuracy of film dosimetry \cite{Paelinck:2007}. This is because this procedure partially accounts for the film response inhomogeneity. Following Ohuchi (2007)\cite{ohuchi:2007} the reflectance can be processed in the same way as the transmittance. Hence, net optical density (NOD)\citep{devic:2005} was defined as:

\begin{equation}
NOD = z = \log_{10} \frac{v_\mathrm{nonirr}}{v_\mathrm{irr}}
\end{equation}

Where $v_{\rm{nonirr}}$ and $v_{\rm{irr}}$ represent pixel values of nonirradiated and irradiated films, respectively, after correction according to lateral correction models.

\subsubsection{Models for lateral correction}

Lateral correction is necessary since the scanner’s response over the scan field is not uniform\cite{Paelinck:2007, fiandra:2006, battum:2008, devic:2006, lynch:2006, Martisikova:2008, Fuss:2007, menegotti:2008, saur:2008} . Deviation from the response in the center of the scanner is particularly important along the {\it x} axis ({\it i.e.}, parallel to the CCD array) and usually negligible along the {\it y} axis ({\it i.e.}, perpendicular to the CCD array). Besides, this correction could also be dependent on the pixel value (or equivalently on the dose). Then, considering that it is negligible along the {\it y} axis, lateral correction is a bidimensional function dependent on PV and pixel position along the {\it x} axis. Different approaches to this correction had been proposed in the literature: lateral correction function has been approximated by a matrix of correction factors \cite{Fuss:2007, Martisikova:2008}, it has been considered independent of the PV \cite{saur:2008}, the dependency on pixel position has been considered a parabola \cite{menegotti:2008}, etc.
Based on the corrections proposed in the literature, four different bidimensional polynomial approximations to the lateral correction function were investigated. All of them are empirical, since the authors did not find in the literature any lateral correction function based on physical assumptions:

Type I:
\begin{equation}
v = a_1 (x-x_c) + a_2 (x-x_c)^{2} + \hat{v} 
\end{equation}

Type II:
\begin{equation}
v = \hat{v} (1 + a_1 (x-x_c) + a_2 (x-x_c)^{2})
\end{equation}

Type III:
\begin{equation}
v = \hat{v} (1 + (a_1 + a_2 \hat{v}) (x-x_c) + (a_3 + a_4 \hat{v}) (x-x_c)^{2})
\end{equation}

Type IV:
\begin{equation}
v = a_1 (x-x_c) + a_2 (x-x_c)^{2} + \hat{v} (1 + a_3 (x-x_c) + a_4 (x-x_c)^{2})
\end{equation}

Here $\hat{v}$ represents "raw" not corrected PV, $x_c$ is the {\it x} coordinate of the center of the scanner, and {\it v} represents corrected PV.

Type I corresponds to an absolute correction independent of dose and second order in the distance from the center. Type II is a relative correction second order in the distance. Type III is a relative correction second order in the distance and in the PV. Type IV is a combination of types I and II.

\subsubsection{Models for sensitometric curves}

Throughout this work, calibration is considered a process that yields the dose measured in a point with pixel position and PVs before and after irradiation given as inputs. Then, it includes lateral correction, NOD calculation and sensitometric curve. Sensitometric curves convert NOD to absolute dose. 

Two types of sensitometric curves with functional forms following the conditions stated by Bouchard \emph{et al} (2009)\cite{bouchard:2009} were studied.  

The first one is an empirical curve, a polynomial fit of order $n$:

\begin{equation}
D = \sum_{i=1}^{n} b_i z^i
\end{equation}

where $z$ represents NOD. 

The second one is based on physical assumptions. A gamma-distributed single hit model derived from percolation theory\cite{moral:2009}:

\begin{equation}
D = \left( \frac{b_1}{b_2 - z}\right)^{1/b_3} - \left( \frac{b_1}{b_2}\right)^{1/b_3}
\end{equation}

\subsection{Calibration with plan-based method}

\subsubsection{Matrix of data}

Increasing the number of dose levels decreases the uncertainty in the sensitometric curve\citep{bouchard:2009}. This is especially important when film response is inhomogeneous. Considering this, a film was irradiated with a $60^{\circ}$ Enhanced Dynamic Wedge (EDW) field of dimensions 20$\times$20 ${\rm cm^2}$ with 438 MU (doses on the film ranging from approximately 75 cGy to approximately 400 cGy). The film was digitized before and after exposure at 150 dpi, obtaining a matrix of data with 1200$\times$1500 px. A margin of 100$\times$200 px was avoided during the computation.

A plan with the geometry of the irradiation was calculated using Eclipse v.10.0 (Varian Medical Systems) treatment planning system (TPS) with anisotropic analytical algorithm (AAA). The accuracy of the calculation of EDW by the TPS was commissioned previously with ionization chamber, linear diode array and 2D ion chamber array. The calculated absolute dose distribution on the plane of the film was exported to a matrix with resolution 0.59 mm/px. It was bilinearly interpolated to register to the film. 

Using the described method, a matrix of 1100$\times$1300 data points was obtained. Each data point included: {\it x} and {\it y} coordinates of the pixel, dose calculated by the TPS and pixel values before and after irradiation for all three color channels (R, G and B).

\subsubsection{Model selection}

To select the most appropriate lateral correction and sensitometric curve model, {\it i.e.}, the calibration model, least squares fitting was used as a maximum likelihood estimation. Therefore, the most probable calibration model is the one that minimizes the root-mean-square error (RMSE) of the differences between the doses measured with film and the doses calculated by the TPS. 

To calculate the RMSE of a calibration model a program was depeloved in C++. A genetic algorithm\cite{Goldberg:1989} searched the parameters that minimize the RMSE of the calibration model. Using a PC with Intel Core 2 Duo 3.0 GHz Processor, the computation time was around 5 min per color channel. Optimized RMSEs for the red channel were obtained for different calibration models.
 
Once a calibration model was selected, all three color channels were calibrated with the genetic algorithm search. After that, film doses were calculated for each color channel. To combine the calibration of all three channels, the weighted mean dose was calculated. Channel doses were weighted with the variance estimated as the square of the RMSE of the channel.

\subsection{Calibration with fragments}

In the literature, the most frequent calibration method uses fragments irradiated with different doses and scanned in different positions over the scanner\cite{bouchard:2009, devic:2005, Martisikova:2008, Fuss:2007}. This method was also followed in this work to compare it with the new plan-based method proposed. 

\subsubsection{Lateral correction}

To evaluate the nonuniformity of the dosimetric system, five film fragments with different PV levels were digitized at different positions on the scanner. One of the fragments was nonirradiated, while the other four had been previously irradiated with different doses. Every fragment was scanned at 18 positions along the {\it x} axis. Two of the fragments were also scanned at 11 positions along the {\it y} axis. A 100$\times$100 px ROI with resolution of 150 dpi was measured at the center of the fragment for every position.  

\subsubsection{Sensitometric curve}

A film was divided in 12 fragments which were irradiated with 25, 50, 100, 150, 175, 200, 225, 250, 300, 350, 400 and 500 MU respectively (104 MU corresponded to a dose of 1 Gy). Before irradiation and 24h after irradiation, each fragment was centered on the scanner and a 100$\times$100 px ROI at 150 dpi was measured at the center of the fragment. 

\subsection{Comparison of calibration methods}

\subsubsection{Intralot and interlot variations}

Three films from lot A03171101A (lot A) were calibrated using the plan-based method. Only the data from the red channel were used. Calibration parameters obtained for the three films from lot A and for the film previously used for plan-based calibration of lot A04141003BB (lot B) were employed to measure the dose on one of the three films from lot A. The RMSE of the differences between the doses measured with film and the doses calculated by the TPS was computed for each set of calibration parameters.

\subsubsection{Verification tests}

Seven different cases were tested. They were based on the IAEA TECDOC-1583 \cite{iaea:1583} tests for commissioning of TPS. They were planned using Eclipse TPS with AAA. Films were posteriorly irradiated in the phantom according to the plans. The geometry of the test cases is described in Table~\ref{tab:Tests}. The images were digitized before and 24 h after irradiation.

\begin{table*}
\caption{\label{tab:Tests} Geometry of the test cases. In all tests a CIRS Thorax phantom (Model 002LFC) was irradiated with an energy of 6 MV in SAD set-up.}
\begin{ruledtabular}
\begin{tabular}{lccccr}
Test & Description & Field size (${\rm cm^2}$) & Gantry angle & Collimator angle & Beam modifiers\\
\hline
1 & Square 10$\times$10 ${\rm cm^2}$ & 10$\times$10 & 0 & 0 &\\ 
2 & Small field & 4$\times$4 & 0 & 0 &\\ 
3 & Lateral incidence & 10$\times$10 & 90 & 0 &\\ 
4 & Tangential field & 10$\times$15 & 90 & 90 &\\ 
5 & Four field box & 10$\times$15 & 0 & 0 &\\ 
&& 8$\times$15 & 90 & 0 &\\ 
&& 10$\times$15 & 180 & 0 &\\ 
&& 8$\times$15 & 270 & 0 &\\ 
6 & EDW and asymmetric fields A & 10$\times$10 & 300 & 90 & EDW15IN\\ 
&& 20$\times$20 & 60 & 90 & EDW30OUT\\
&& 40$\times$40 & 180 & 90 &\\
&& 10$\times$15 & 0 & 90 & \\ 
7 & EDW and asymmetric fields B & 10$\times$10 & 300 & 90 & EDW45IN\\ 
&& 20$\times$20 & 60 & 90 & EDW60OUT\\
&& 5$\times$5 & 180 & 90 &\\
&& 15$\times$15 & 0 & 90 & \\ 
\end{tabular}
\end{ruledtabular}
\end{table*}

The calculated dose distribution on the plane of the film was exported with resolution 0.59 mm/px. The digitized films were converted to dose according to the previously selected calibration model. Four sets of images were created based on the parameters derived from: fragments using only the red channel, fragments using the three channels, plan-based red channel and plan-based three channels. 2D gamma analysis of the test cases was conducted. The selected criteria for the analysis were 4 \% 3 mm excluding points with less than 20 \% of the maximum dose. To automate the procedure of dose calculation and gamma analysis, a program was developed in C++.

\section{Results and discussion}

\subsection{Preliminary tests}

\subsubsection{Effect of surrounding film}

\begin{figure}
\includegraphics[width=\linewidth]{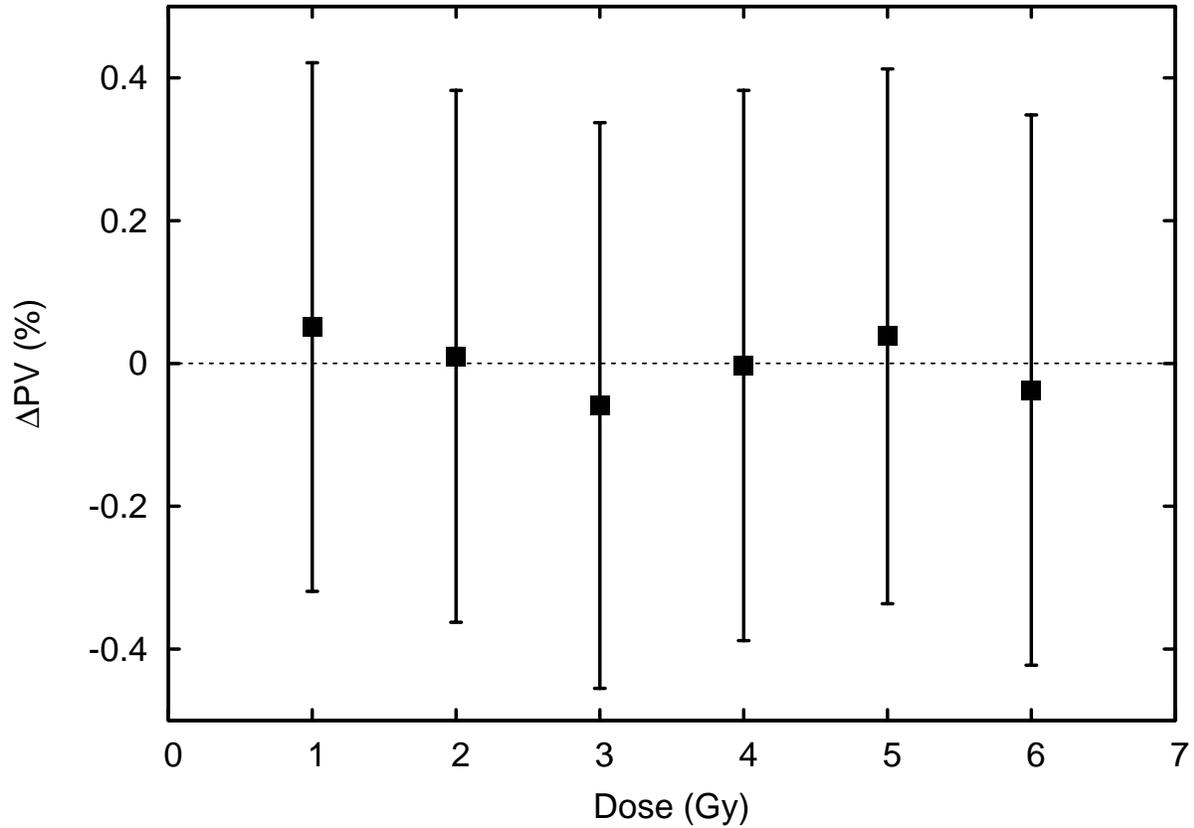}
\caption{\label{fig:surrounding} Mean and standard deviation of the PV measured in a nonirradiated fragment as a function of the dose of an abutting fragment, data are scaled so that the y-axis represents relative deviations of the PV with respect to the average of all six measurements.
}\end{figure}

Contrary to Richley \emph{et al} (2010)\citep{Richley:2010}, no effect of surrounding film was found in this study. Fig.~\ref{fig:surrounding} shows mean and standard deviation of the PV of the red channel measured in the nonirradiated fragment as a function of the dose of an abutting fragment, data are scaled so that the y-axis represents relative deviations of the PV with respect to the average of all six measurements. No statistically significant (linear) correlation between PV of a fragment and dose of an abutting fragment was found (p = 0.46).  

A slight effect of surrounding film is to be expected in the immediate vicinity of a stepwise change in dose, since the point spread function of the system cannot be a Dirac delta function. However, if the effect is significant some milimeters away from the step, the digitized image should be blurred. Possible explanations for the effect of surrounding film found by Richley \emph{et al} could be a problem with the optics of the scanner in reflection mode or a variation in the temperature of the scanner's bed.   

Since no effect of surrounding film was found, this opens the possibility of using the same calibration methods in reflection mode as in transmission mode with the dosimetric system formed by Gafchromic EBT2 films and Epson Expression 10000XL scanner. 
 
\subsubsection{Film response homogeneity}

The contribution of film response inhomogeneity to the uncertainty of the measured dose is known to be substantial\cite{Martisikova:2008}. It is reduced if the film is digitized before irradiation and NOD is calculated. However, this only accounts for the background PV or fog, and not for differences in sensitivity (e.g. due to thickness variation of the active layer).

\begin{figure*}
\begin{minipage}[b]{0.47\linewidth}
\centering
\includegraphics[width=\linewidth]{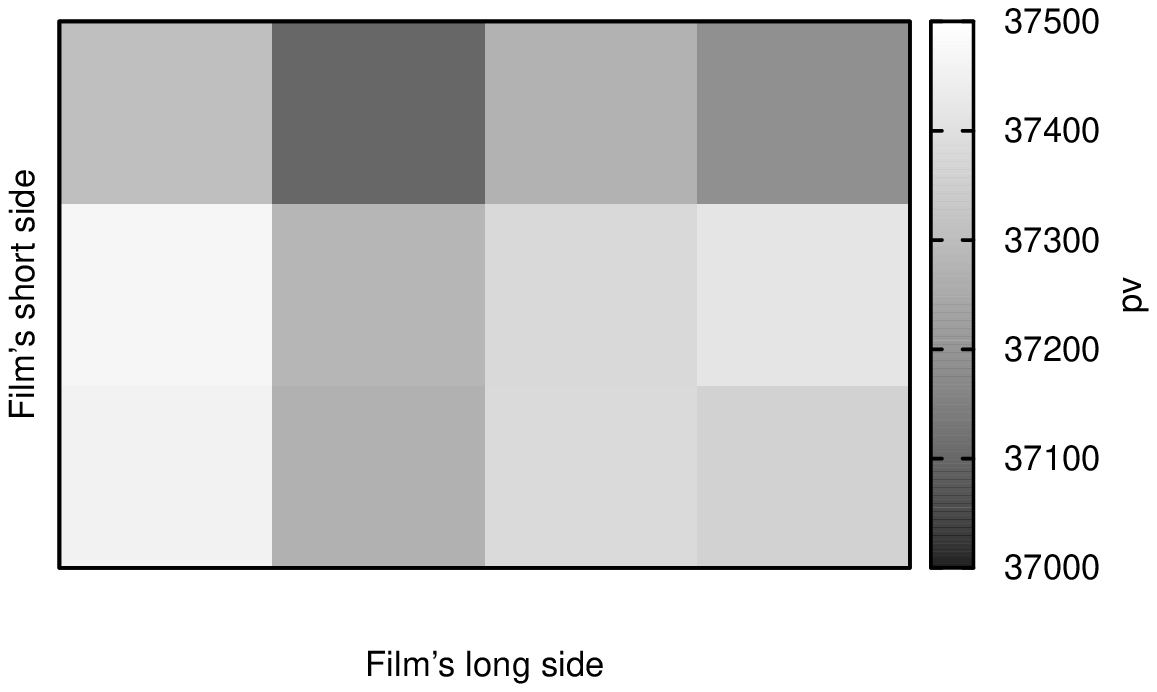}\\
(a)
\end{minipage}
\hfill
\begin{minipage}[b]{0.47\linewidth}
\centering
\includegraphics[width=\linewidth]{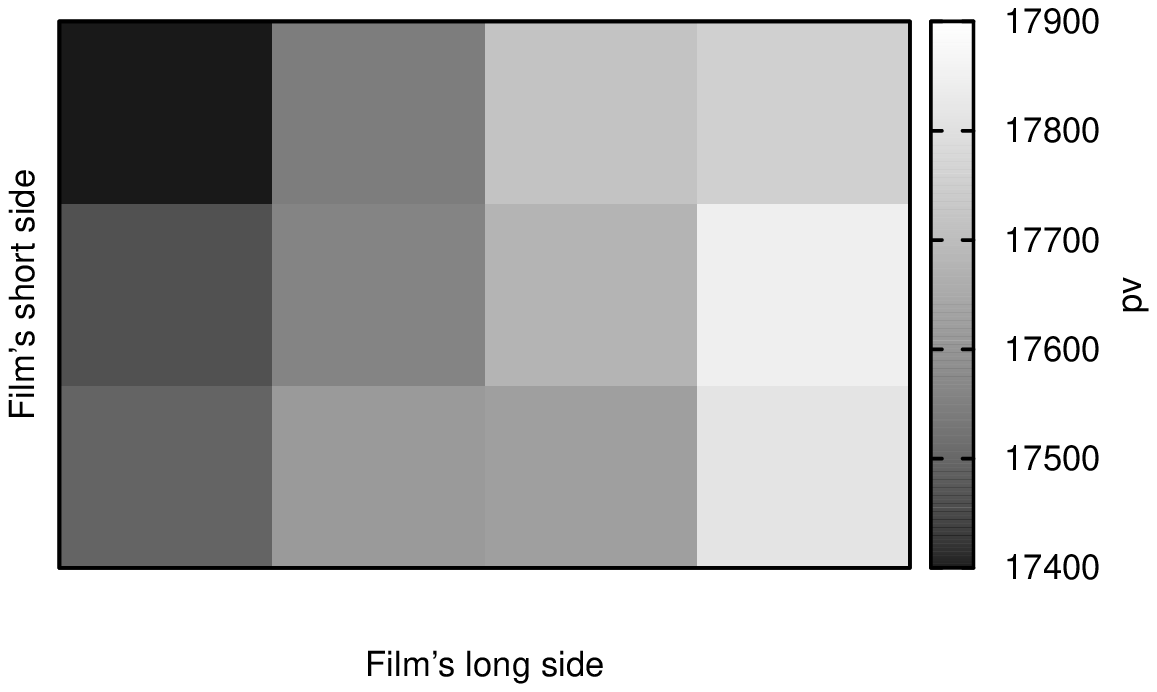}\\
(b)
\end{minipage}
\caption{\label{fig:homogeneity} Mean PVs measured on different fragments of a film (a) nonirradiated and (b) irradiated with 2 Gy. Every fragment was centered in the scan and a 100$\times$100 px ROI was measured at the center of the fragment.}\end{figure*}

Fig.~\ref{fig:homogeneity} presents mean PVs of the red channel measured on different fragments of a film nonirradiated and irradiated with 2 Gy. One-way analysis of variance (ANOVA) found statistically significant differences between PVs measured on different fragments for both nonirradiated (p $<$ 0.001) and irradiated films (p $<$ 0.001). Therefore, film response inhomogeneity is significant in Gafchromic EBT2 films. To check if this inhomogeneity is only due to fog differences, NOD was calculated for each pixel in the film. One-way ANOVA found statistically significant differences between fragments (p $<$ 0.001). Maximum differences of 0.01 NOD between mean NODs of fragments irradiated with 2 Gy were observed with Tukey's HSD test.   

Hence, film response inhomogeneity in Gafchromic EBT2 films increases the uncertainty of the dosimetry affecting the calibration and the final results. The use of NOD alone cannot correct this defect. More advanced procedures to correct film response inhomogeneity are necessary.
   
\subsection{Calibration with plan-based method}

\subsubsection{Model selection:}

Optimized RMSEs for the red channel obtained with different calibration models are shown in Table~\ref{tab:rmse}. With respect to lateral correction, type I and type IV lateral correction functions obtained the lowest RMSE of 4.4 cGy, whereas type III obtained 4.5 cGy and type II 9.9 cGy. Considering that the sample size contains a matrix of 1100$\times$1300 data points, these differences suggest substantial evidence for the superiority of models type I and type IV according to the Akaike information criterion (AIC). Since type IV includes type I functions, this supports the conclusions of Saur \emph{et al}, (2008)\cite{saur:2008} who found a better agreement with absolute corrections independent of dose. Therefore, type I lateral correction functions were selected.

With respect to sensitometric curves, gamma-distributed single hit model and polynomial fits of orders 4 and 5 had the same RMSE. The complexity of the model was considered negligible comparing the number of parameters with the number of data points to fit. However less complexity facilitates the optimization. Gamma-distributed single hit curves exhibited less robustness during the optimization, {\it i.e.} small changes of the algorithm affected considerably the minimum RMSE found. Therefore, polynomial fit of order 4 was selected as the sensitometric curve.  

It is important to note that even though genetic algorithm is a well established optimization method that effectively escapes from local minima, its results cannot be taken as global minima. As a consequence, it should not be concluded that the model selected is the best of the models analyzed, although this hypothesis becomes more plausible.

\begin{table}
\caption{\label{tab:rmse} Optimized RMSE for the red channel obtained with different calibration models.}
\begin{ruledtabular}
\begin{tabular}{lcr}
Lateral correction&Sensitometric curve&RMSE (cGy)\\
\hline
Type I & Polynomial order 4 & 4.4\\
Type II & Polynomial order 4 & 9.9\\
Type III & Polynomial order 4 & 4.5\\
Type IV & Polynomial order 4 & 4.4\\
Type I & Polynomial order 3 & 4.5\\
Type I & Polynomial order 5 & 4.4\\
Type I & Single hit model & 4.4\\
\end{tabular}
\end{ruledtabular}
\end{table}

\subsection{Calibration with fragments}

\subsubsection{Lateral correction}

\begin{figure*}
\begin{minipage}[b]{0.47\linewidth}
\centering
\includegraphics[width=\linewidth]{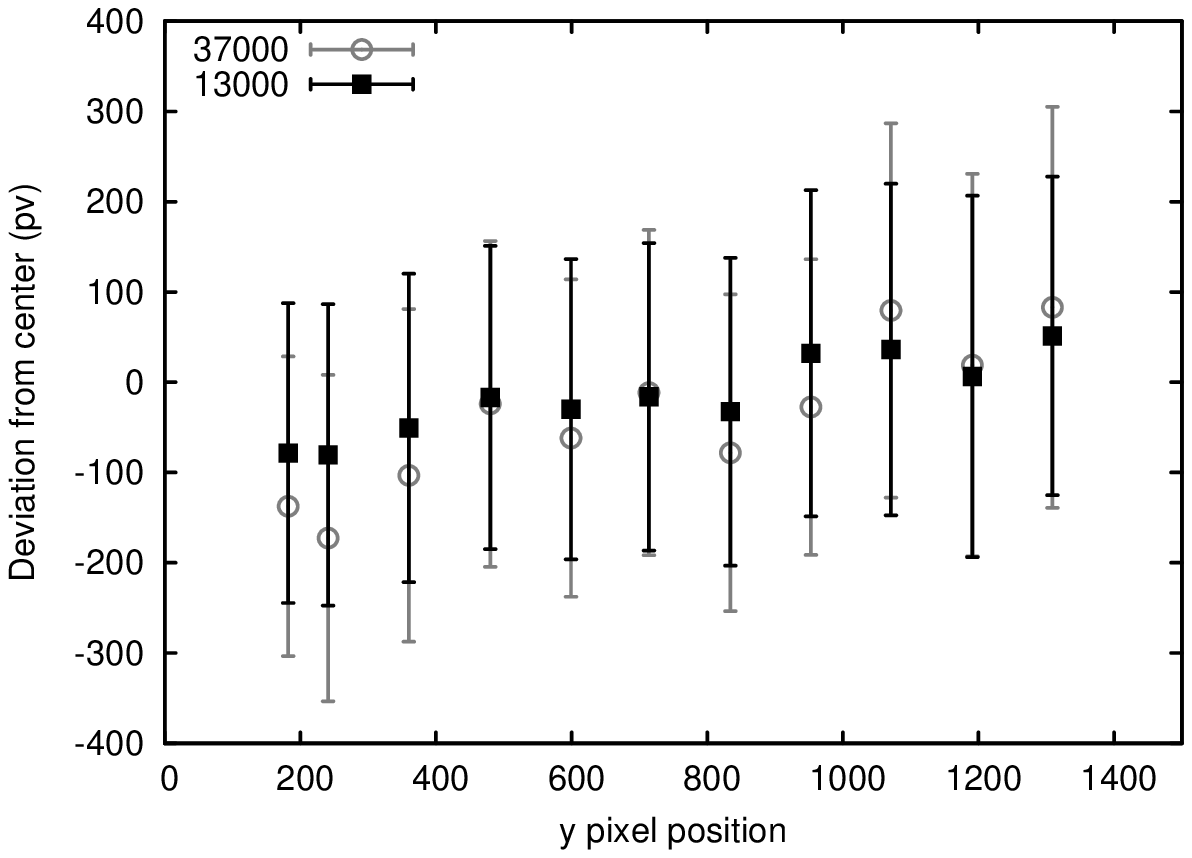}\\
(a)
\end{minipage}
\hfill
\begin{minipage}[b]{0.47\linewidth}
\centering
\includegraphics[width=\linewidth]{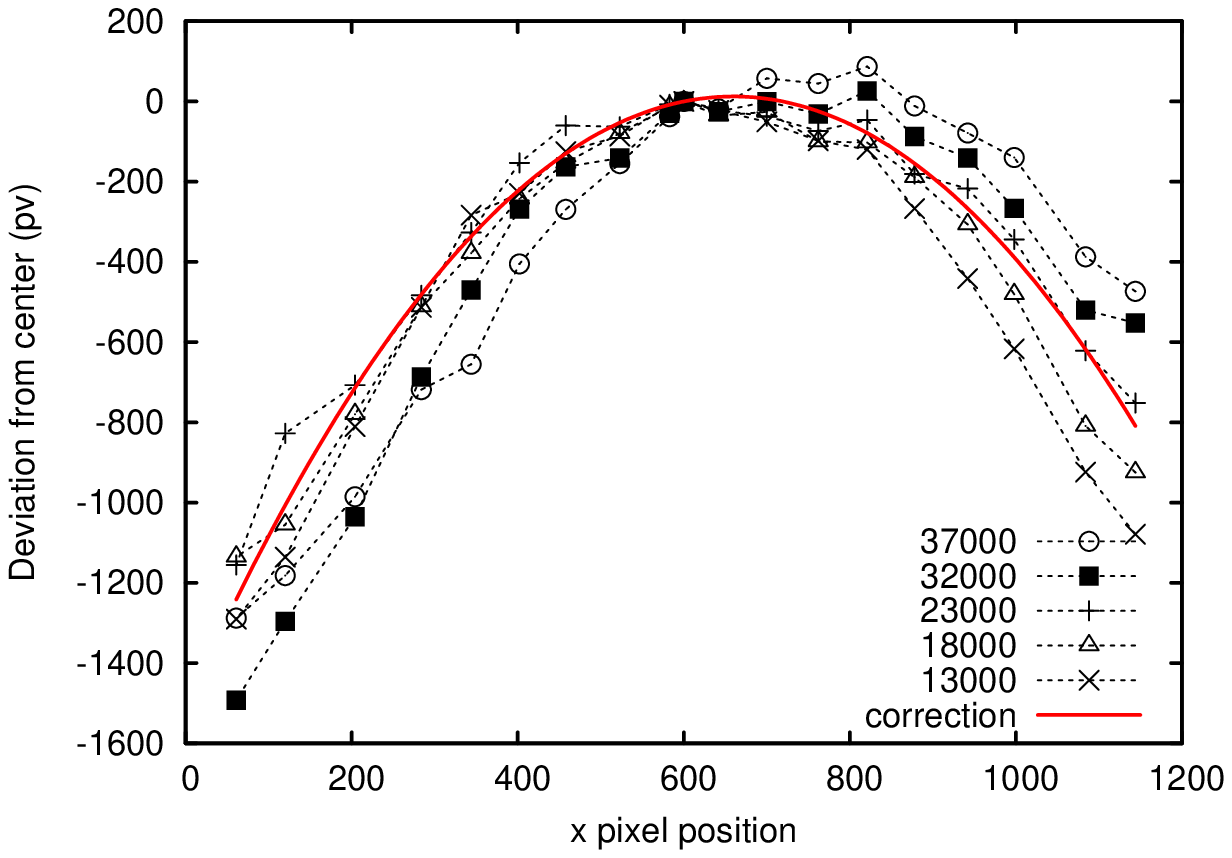}\\
(b)
\end{minipage}
\caption{\label{fig:lat_correction}Deviation from the value in the center of the scanner for different pixel value levels, as a function of the pixel position (a) along the {\it y} axis ({\it i.e.}, perpendicular to the CCD array) (b) along the {\it x} axis ({\it i.e.}, parallel to the CCD array); the fixed line represents the lateral correction fitted from the fragments.}
\end{figure*}

Deviation from the value in the center of the scanner for different PV levels in the red channel, as a function of the pixel position along the {\it y} axis and along the {\it x} axis is illustrated in Fig.~\ref{fig:lat_correction}. One-way analysis of variance (ANOVA) found statistically significant differences between PVs as a function of the pixel position along the y axis (p $<$ 0.001) and along the x axis (p $<$ 0.001). A linear regression of the measurements along the y axis obtained maximum differences of 0.002 NOD for doses around 4 Gy. Hence, along the y axis the lateral correction was considered to be negligible. Along the x axis the nonuniform response was fitted according to the model selected. The lateral correction function fitted from the fragments is shown.

\subsubsection{Sensitometric curve}

\begin{figure}
\includegraphics[width=\linewidth]{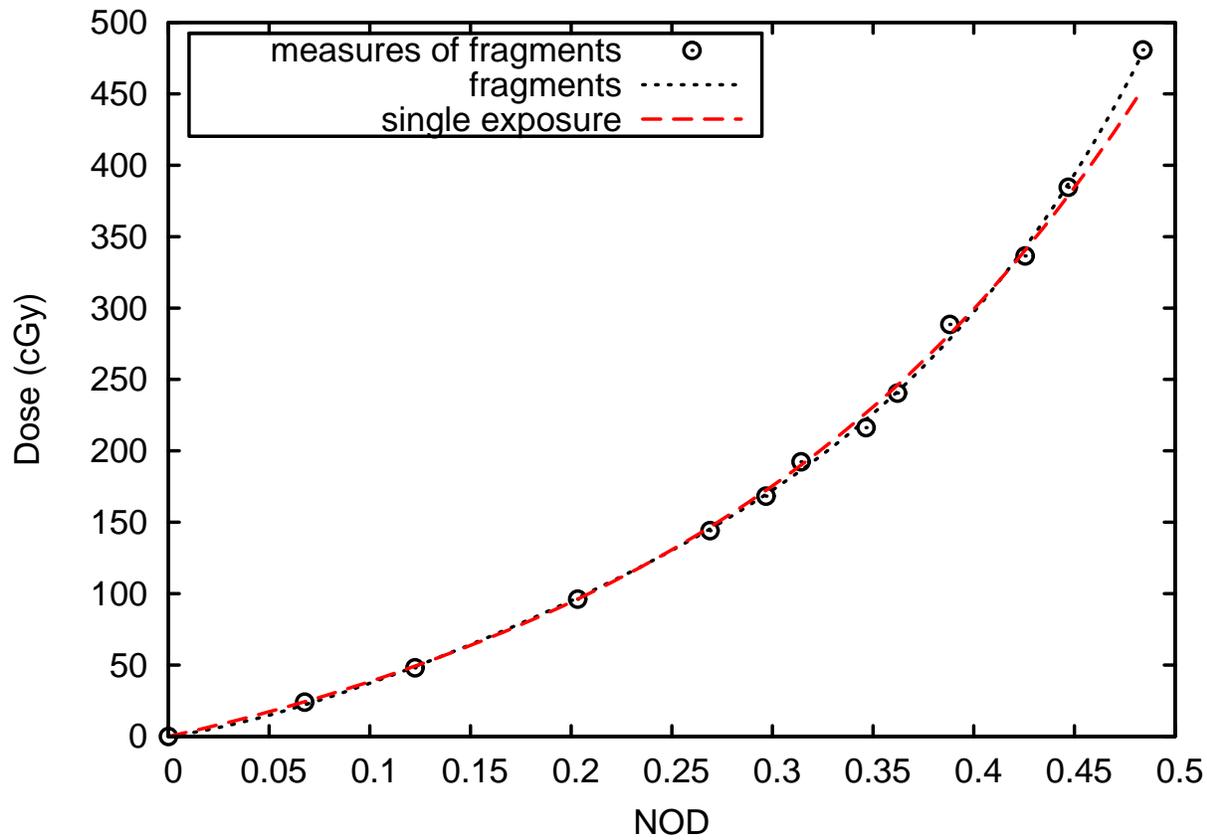}
\caption{\label{fig:fragments} Sensitometric curves obtained with fragments (dotted) and with plan-based method (dashed). The sensitometric curve obtained with plan-based method is extrapolated for doses greater than 400 cGy.}\end{figure}

The sensitometric curve fitted from the fragments for the red channel, according to the model selected, is shown in Fig.~\ref{fig:fragments}. It is compared with the sensitometric curve obtained for the red channel with the plan-based method. The sensitometric curve obtained with the plan-based method is extrapolated for doses greater than 400 cGy. 

\subsection{Comparison of calibration methods}

Table~\ref{tab:compara} compares RMSEs obtained with different calibration methods and calculated on different films. In the first part, plan-based calibration methods are compared. According to the AIC, the RMSEs present substantial evidence for the superiority of three-channel plan-based dosimetry compared to one-channel plan-based dosimetry. This is also the case for calibration with fragments, as it is shown in the second part. It has to be noted that pixel measures were aggregated in every ROI for the calculation of RMSEs on the sensitometric curve from fragments. These RMSEs would increase if pixel measures were disaggregated, and even more so if not only the sensitometric curve but also the residuals of the lateral correction's fit were considered. To calculate the weighted mean dose, channel doses were weighted with the variance estimated as the square of the RMSE of the channel. In the third part, calibration parameters obtained from calibration with fragments were employed to calculate RMSE on the film used for plan-based calibration. Calibration with fragments showed worse RMSEs than plan-based calibration. This outcome could be partially explained by the fact that the calibration with fragments is less robust to film inhomogeneities than plan-based calibration: plan-based calibration can use every pixel of the film, whereas calibration with fragments only uses a limited number of pixels which share coordinates. Calibration with fragments also needs a more complex measuring process. In addition, film-to-film variations are an important source of uncertainty too, as it is showed in Table~\ref{tab:intralot}. 

\begin{table*}
\caption{\label{tab:compara} RMSEs obtained with different calibration methods and calculated on different films. Pixel measures were aggregated in every ROI for the calculation of RMSEs on the sensitometric curve from fragments.}
\begin{ruledtabular}
\begin{tabular}{lccr}
Calibration method & Sample & Color channel & RMSE (cGy)\\
\hline
Plan-based & Plan-based film & Red & 4.4\\
& & Green & 3.9\\
& & Blue & 10.3\\ 
& & 3 channel weighted mean & 3.8\\ 
\hline
Fragments & Sensitometric curve from fragments & Red & 5.0\\ 
& & Green & 4.5\\ 
& &  Blue & 13.4\\ 
& & 3 channel weighted mean & 4.4\\ 
\hline
Fragments & Plan-based film & 3 channel weighted mean & 7.3\\ 
\end{tabular}
\end{ruledtabular}
\end{table*}

\subsubsection{Intralot and interlot variations}

Table~\ref{tab:intralot} shows the influence of intralot (film-to-film) and interlot variations on RMSEs. The RMSEs were calculated on film 1 (Lot A) with calibration parameters derived from different films of two different lots (A and B). Film-to-film variations were found not negligible. This implies that to decrease the uncertainty of the sensitometric curve could be necessary not only to increase the number of dose levels, but also to calibrate several films simultaneously. Considering this, a possible improvement for the plan-based calibration method presented in this work would be to optimize simultaneously several films irradiated according to one or more reference plans.

\begin{table}
\caption{\label{tab:intralot} Intralot and interlot variations. RMSEs were calculated on film 1 (Lot A) with calibration parameters derived from different films of two different lots.}
\begin{ruledtabular}
\begin{tabular}{lcr}
Test & Calibration on film & RMSE (cGy)\\
\hline
Intralot variation & Film 1 (Lot A) & 5.8\\
& Film 2 (Lot A) & 7.3\\
& Film 3 (Lot A) & 6.3\\
\hline
Interlot variation & Film 1 (Lot B) & 16.0\\
\end{tabular}
\end{ruledtabular}
\end{table}

\subsubsection{Verification tests}

\begin{figure}
\includegraphics[width=\linewidth]{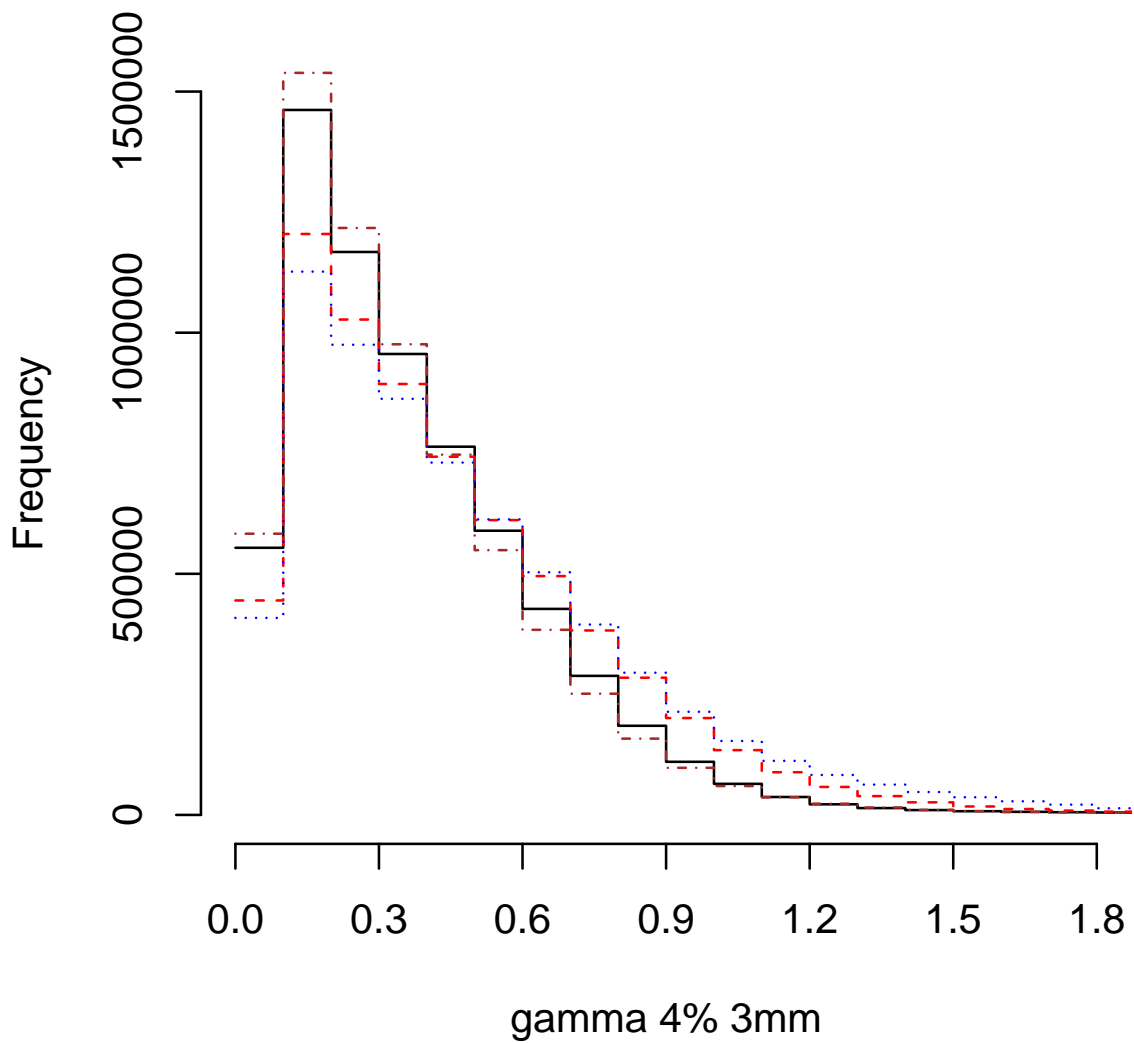}
\caption{\label{fig:verification} Histograms of gamma (4\% 3mm) values obtained with different calibration methods: fragments red channel (dotted), fragments 3 channels (dotdash), plan-based red channel (dashed) and plan-based 3 channels (solid). Gamma values were calculated for all the points in the test cases excluding points with less than 20 \% of the maximum dose of the test.}\end{figure}

In Fig.~\ref{fig:verification}, histograms of gamma (4\% 3mm) values obtained with different calibration methods are plotted. Gamma values were calculated for all the points in the test cases excluding points with less than 20 \% of the maximum dose of the test. In Table~\ref{tab:Gamma}, the percentage of points with $\gamma_{<1}$ and the $\gamma_{\rm{mean}}$ calculated with the compared calibration methods are presented for each test case, as well as the average values calculated for all the points in the test cases. 

Three-channel calibration methods showed the best agreement with the TPS, followed by red channel plan-based; red channel calibration with fragments showed the worst agreement. The average number of points with $\gamma_{<1}$ was 90.7 \% with red channel fragments, 93.3 \% with red channel plan-based and 96.6 \% with both three-channel plan-based calibration and three-channel calibration with fragments. The average $\gamma_{\rm{mean}}$ was 0.49 with red channel calibration with fragments, 0.46 with red channel plan-based, 0.39 with three-channel plan-based calibration and 0.38 with three-channel calibration with fragments.

The plan-based calibration method obtained comparable results to the well-established calibration method with fragments. It indicates that the plan-based calibration method is a feasible alternative to the calibration with fragments. However, possible film-to-film variations or systematic inaccuracies of the TPS cannot be excluded. The plan-based calibration method offers increased robustness against film response inhomogeneities (since it can use every pixel of the film) and reduces considerably the time required for calibration (in this work, calibration time was reduced from several hours for calibration with fragments to minutes for plan-based calibration).

\begin{table*}
\caption{\label{tab:Gamma} 2D gamma analysis (4\% 3mm) of the test cases with different calibration methods.}
\begin{ruledtabular}
\begin{tabular}{lcccccccc}
Test & \multicolumn{4}{c}{$\gamma_{<1}$ (\%)} & \multicolumn{4}{c}{$\gamma_{mean}$} \\ \cline{2-5} \cline{6-9}
& \multicolumn{2}{c}{Fragments} & \multicolumn{2}{c}{Plan-based} & \multicolumn{2}{c}{Fragments} & \multicolumn{2}{c}{Plan-based}\\ 
 & Red & 3 channels & Red & 3 channels & Red & 3 channels & Red & 3 channels\\
\hline
1 & 92.4 & 94.2 & 88.3 & 92.6 & 0.46 & 0.41 & 0.51 & 0.47\\
2 & 90.3 & 88.6 & 89.2 & 88.8 & 0.47 & 0.51 & 0.50 & 0.50\\
3 & 97.2 & 98.7 & 98.6 & 98.8 & 0.40 & 0.34 & 0.34 & 0.32\\
4 & 93.5 & 95.9 & 97.1 & 97.0 & 0.46 & 0.41 & 0.39 & 0.38\\
5 & 89.1 & 97.2 & 94.7 & 97.9 & 0.46 & 0.32 & 0.42 & 0.34\\
6 & 89.4 & 96.9 & 95.6 & 97.8 & 0.53 & 0.38 & 0.44 & 0.36\\
7 & 87.3 & 97.9 & 87.5 & 96.7 & 0.58 & 0.37 & 0.59 & 0.41\\ \hline
Average result & 90.7 & 96.6 & 93.3 & 96.6 & 0.49  & 0.38 & 0.46 & 0.39\\
\end{tabular}
\end{ruledtabular}
\end{table*}

\section{Conclusions}

Radiochromic dosimetry in reflection mode using Gafchromic EBT2 films was found to be a viable alternative to transmission mode. In this study, no effect of surrounding film was found with the dosimetric system formed by Gafchromic EBT2 films and Epson Expression 10000XL scanner. This opens the possibility of using the same calibration methods in reflection mode as in transmission mode. 

Film response inhomogeneity with EBT2 films was found to be important, not only due to differences in the fog but also to differences in sensitivity. The use of NOD alone cannot correct this defect. More advanced procedures to correct film response inhomogeneity are necessary.

To offer increased robustness against film response inhomogeneities, a novel plan-based calibration method was developed. Plan-­based calibration is a single exposure method that can use every pixel and each of the three color channels of the scanned film to obtain the parameters of the calibration model. Plan-based calibration uses a reference plan (in this study a field with $60^{\circ}$ EDW) calculated by the TPS. The accuracy of the calculation was carefully commissioned. A film was irradiated following the reference plan. Least squares fitting was employed to find the parameters of the calibration model that minimized the differences between TPS and the doses measured with film. The complexity of the optimization made it necessary to use a genetic algorithm search. The calibration model (lateral correction and sensitometric curve models) was selected based on a maximum likelihood analysis. The best results for lateral corrections were obtained using absolute corrections independent of dose. With respect to sensitometric curves, an empirical polynomial fit of order 4 was found to obtain results equivalent to a gamma-distributed single hit model based on physical assumptions. Film-to-film variations were found to be not negligible, thus a possible improvement for the plan-based calibration method presented in this work would be to optimize simultaneously several films irradiated according to one or more reference plans.

Three-channel dosimetry was calculated using the weighted mean dose of the color channels. The variances of the calibration, estimated as the square of the RMSE for each channel, were used as weights. Three-channel dosimetry was found to be substantially superior to red-channel dosimetry. 

Plan-based calibration method was found to be a feasible alternative to the well-established calibration method with fragments. This novel method offers increased robustness against film response inhomogeneities (since it can use every pixel of the film) and reduces considerably the time required for calibration (in this work, calibration time was reduced from several hours for calibration with fragments to minutes for plan-based calibration).  

\begin{acknowledgments}
The authors would like to thank Sebasti{\`a} Agramunt, Primo\v{z} Peterlin, Juanjo Rovira and Attila \v{S}arvari for many useful discussions and contributions to this work.
\end{acknowledgments}

\providecommand{\noopsort}[1]{}\providecommand{\singleletter}[1]{#1}%
\end{document}